\documentclass[
  aps,
  prx,
  reprint,
  superscriptaddress,
  longbibliography
]{revtex4-2}
\usepackage{graphicx}
\usepackage{amsmath,amssymb,bm}
\usepackage{hyperref}
\hypersetup{
  colorlinks=true,
  citecolor=blue,
  urlcolor=blue,
  linkcolor=blue,
  pdfborder={0 0 0}
}
\usepackage{mathrsfs}
\usepackage{physics}
\usepackage{booktabs}
\usepackage{braket}
\usepackage{xcolor}
\usepackage{bibunits}
\defaultbibliographystyle{apsrev4-2}
\defaultbibliography{references}

\begin{document}

\title{Constant-Depth Multi-Product Formula for Trotter Error Mitigation \\in Near-Term Digital Quantum Simulation}
\author{Chan Bin Bark}
\altaffiliation{Electronic Address}
\email{qkrcksqls135@hanyang.ac.kr}
\affiliation{Department of Physics, Hanyang University, Seoul, Republic of Korea}

\author{Sangjin Lee}
\affiliation{Quantum Universe Center, Korea Institute for Advanced Study, Seoul 02455, South Korea}

\author{Kwang Jun Ahn}
\affiliation{Department of Energy Systems
Research, Ajou University,
Suwon 16499, Republic of Korea}

\author{Sangmo Cheon}
\affiliation{Department of Physics, Hanyang University, Seoul, Republic of Korea}

\author{Moon Jip Park}
\altaffiliation{Electronic Address}
\email{moonjippark@hanyang.ac.kr}
\affiliation{Department of Physics, Hanyang University, Seoul, Republic of Korea}

\author{Youngseok Kim}
\altaffiliation{Electronic Address}
\email{kim.10017@osu.edu}
\affiliation{Department of Electrical and Computer Engineering, The Ohio State University, Columbus OH 43210, USA}

\date{\today}

\begin{abstract}
Digital quantum simulation of many-body dynamics faces a tension between algorithmic Trotter error and physical noise that accumulates with circuit depth. Typical higher-order product formulas mitigate the algorithmic error at the cost of deeper circuits. Our benchmark shows that, within the limited physical error budget, the feasible advantage is confined to absolute errors well below the $10^{-2}$ scale, which vanishes under noise levels of current quantum hardware. In sharp contrast to the previous Trotter error mitigation methods, we introduce a constant-depth multi-product formula (cd-MPF) that suppresses the Trotter error by combining multiple circuits at fixed circuit depth. We identify an auxiliary parameter $\alpha$, which reshapes the Trotter error terms while leaving the target evolution invariant. The classical linear combination of the measured expectation values cancels the leading $(\Delta t)^{2}$ algorithmic contribution and steepens the Trotter-error scaling with the two-qubit circuit depth $d$ from $d^{-2}$ to $d^{-4}$. Combined with physical-noise mitigation, our method can serve as a key ingredient for realizing long-time quantum dynamics simulation on near-term hardware.
\end{abstract}

\maketitle

\begin{bibunit}[apsrev4-2]

\section{Introduction}\label{sec:intro_overview}

Efficient simulation of many-body quantum systems is one of the central goals in quantum information science. Within this framework, modeling quantum dynamics is uniquely promising since a quantum device can natively encode the multi-particle entanglement and superposition of the system. This capability makes the real-time evolution of many-body Hamiltonians a natural target for digital processors, as it governs diverse non-equilibrium phenomena ranging from molecular reaction dynamics to transport in correlated matter~\cite{Georgescu2014, Daley2022}.

Along this line, digital quantum processors implement continuous time evolution through a universal gate set and can therefore, in principle, target arbitrary Hamiltonians on a single hardware platform. The price for this generality is that the continuous time-evolution operator $e^{-iHT}$ must be decomposed into a sequence of elementary gates, most commonly through a Trotter--Suzuki product formula~\cite{Trotter1959, Suzuki1976, Lloyd1996, Georgescu2014}. This Trotterization introduces an unavoidable algorithmic error whose magnitude depends on the Trotter step size, the order of the product formula, and the algebraic structure of the non-commuting Hamiltonian terms.

Considerable effort has been devoted to controlling and suppressing this algorithmic error. Tight error bounds with commutator scaling have clarified how product-formula errors depend on the structure of the Hamiltonian and have guided the design of improved decompositions~\cite{Childs2021, Childs2018, Heyl2019, Lane2025}. Higher-order product formulas (HOPFs) that suppress the algorithmic error by tuning the ordering or the weights of the elementary gates have also been developed~\cite{suzuki1990, yoshida1990, Berry2007, Wiebe2010}, and beyond a single product formula, multi-product formulas (MPFs) take a weighted linear combination of expectation values obtained from product-formula circuits with different step sizes and recover a higher effective order of accuracy at the level of estimated observables~\cite{CarreraVazquez2023, Zhuk2024, Robertson2025}. Both HOPFs and MPFs have been shown to reduce the algorithmic error substantially in idealized noiseless settings. However, both strategies generally enlarge the per-step gate count or require longer circuits to be executed on hardware, which directly amplifies the impact of physical noise on present-day devices.

The role of physical noise has been brought into sharp focus by recent hardware demonstrations of quantum dynamics on noisy intermediate-scale quantum processors~\cite{Smith2019}. Furthermore, recent utility-scale simulations of the dynamics of the transverse-field Ising 
were informative only when the error accumulated along the circuit was actively suppressed~\cite{Kim2023, haghshenas2026digital}, showing that each additional gate layer spent on refining the algorithmic accuracy must
be paid for in physical error. This recognition has motivated Trotter-error mitigation schemes that suppress the algorithmic error
through classical post-processing rather than deeper circuits. Representative examples include Richardson extrapolation of Trotterized expectation values~\cite{Endo2019}, polynomial-interpolation refinements~\cite{Rendon2024, Watson2025}, and classical combinations of short circuits~\cite{Lee2025}.

Despite this progress, the joint trade-off between algorithmic and physical errors in the regime that is most relevant to near-term experiments is still not well understood. Under realistic physical error, do conventional depth-increasing strategies such as HOPFs still pay off? Or can we instead mitigate the Trotter error at fixed depth, keeping the circuit (physical) noise unchanged? In this work we address these questions systematically. We benchmark several HOPFs against the depth-matched second-order Leapfrog formula, develop a constant-depth Trotter-error mitigation scheme based on a tunable Trotter parameter, and analyze how algorithmic and physical errors combine to set the optimal operating point of a noisy digital quantum simulation.

This work is structured as follows. Section~\ref{sec:PF} reviews product formulas and the observable Trotter error, and
Sec.~\ref{sec:tradeoff} formulates the trade-off between the Trotter error and physical noise, establishing the two-qubit circuit depth as a proxy for the physical-noise budget. Section~\ref{sec:mitigation} introduces the cd-MPF, which cancels the leading Trotter error through a classical combination of circuits sharing the same depth.
Section~\ref{sec:comparison} presents the depth-matched benchmarks against the HOPFs, and Sec.~\ref{sec:noise} examines the cd-MPF under an explicit depolarizing-noise model. We conclude with a discussion of the operating range and possible extensions of constant-depth mitigation.

\section{Higher-order product formula and Trotter error}\label{sec:PF}
To simulate many-body quantum time evolution on digital quantum computers via Trotterization, the system Hamiltonian is decomposed into a sum of $K$ non-commuting sub-Hamiltonians,
\begin{align}
H=\sum_{\mu=1}^{K}H_\mu ,
\end{align}
and the target time evolution is described by the time-evolution operator, $U(T)=e^{-iHT}$. Since $U(T)$ cannot be directly implemented as a native operation on digital quantum computers, it is approximated by an ordered sequence of the elementary sub-evolutions, $U_\mu(t)=e^{-iH_\mu t}$. For non-commuting sub-Hamiltonians $([H_\mu, H_\nu] \neq 0$ for $\mu\neq \nu \in \{1,\cdots ,K\})$, a general $n$-th order product formula decomposes the full dynamics as a product of these elementary operators,
\begin{align}
U(T)=e^{-iHT} \simeq \left[ \prod_{j=1}^{M} \exp\left( -i a_j H_{\nu_j}\Delta t \right) \right]^r .
\label{eq:gpf}
\end{align}
Here, $r$ denotes the number of Trotter steps, and each step corresponds to a time evolution of the discrete time interval $\Delta t = T/r$. The integer $M$ is the number of elementary evolution operators in a single step. The coefficients $a_j$ determine the effective evolution time of each sub-Hamiltonian within one time step, while $\nu_j\in\{1,\dots,K\}$ specifies which sub-Hamiltonian is applied at the $j$-th position in the sequence. 

Generally, HOPFs are constructed by choosing different sets of $\{a_j,\nu_j\}$ in Eq.~\eqref{eq:gpf}. Moreover, a palindromic (symmetric) decomposition $U(\Delta t)$, satisfying
$U(-\Delta t) = \left[U(\Delta t)\right]^{-1}$, contains only even-order Trotter errors in $\Delta t$. For two non-commuting Hamiltonians $H_A$ and $H_B$$(H=H_A + H_B)$, the simplest symmetric Trotter decomposition is the \emph{Leapfrog formula},
\begin{align}
U_{\mathrm{LF}}(\Delta t) = e^{-iH_A \Delta t/2}e^{-iH_B \Delta t}e^{-iH_A \Delta t/2},
\end{align}
whose palindromic structure makes it second-order accurate: the leading observable error is $O(\Delta t^2)$ instead of $O(\Delta t)$, i.e. $n=2$. Although HOPFs can in principle reduce the algorithmic error further, we adopt the Leapfrog formula as the baseline of the benchmark for the analysis that follows.

% The Lie--Trotter formula, as the simplest example, is given by
% \begin{align}
% U_{\mathrm{LT}}(\Delta t) = \prod_{\mu=1}^{K} e^{-iH_\mu \Delta t},
% \end{align}
% which corresponds to the specific choice of $M = K$, $a_j = 1$, and $\nu_j = j$. 

When simulating the Hamiltonian up to time $T$, the algorithmic error induced by the Trotterized time-evolution operator can be expanded in powers of the time step $\Delta t$ for a given observable $O$ as
\begin{align}
|\langle O\rangle_{\mathrm{exact}} - \langle O\rangle_{\mathrm{Trotter}}| = \sum_{j=2}^{\infty} \omega_{j}\, (\Delta t)^j ,
\label{eq:obserrorexpansion}
\end{align}
 where $\langle O\rangle_{\mathrm{exact}}$ and $\langle O\rangle_{\mathrm{Trotter}}$ denote the expectation values obtained after the exact and Trotterized time evolutions, respectively. The scalar coefficient $\omega_j$ represents the error contribution at order $j$, which is fundamentally governed by combinations of nested commutators among the non-commuting sub-Hamiltonians. We note that the construction of a product formula with a given leading-order accuracy is not unique; even when two formulas share the same formal order, the structure and magnitude of the remaining higher-order error coefficients can differ depending on the operator ordering and the choice of coefficients $a_j$. The actual Trotter error observed in a physical simulation therefore depends sensitively on the detailed decomposition, the observable $O$, and the initial state, even when the nominal leading-order scaling is fixed.

\begin{figure}[t]
\centering
\includegraphics[width=\columnwidth]{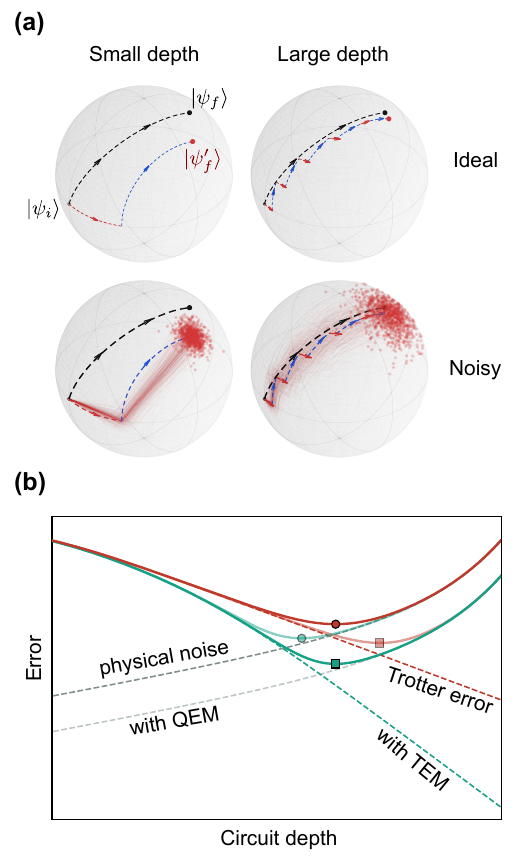}
\caption{\textbf{Trade-off between Trotter error and physical noise in Trotterized quantum simulation.} \textbf{(a)} Bloch-sphere illustration of ideal and noisy Trotterized time evolutions. The black dashed curve denotes the exact trajectory, and the red and blue dashed curves denote partial evolutions generated by different sub-Hamiltonians. The exact and Trotterized final states, $\ket{\psi_f}$ and $\ket{\psi_f'}$, are shown as black and red points, respectively. The upper panels show ideal evolutions with 1 and 5 Trotter steps; the lower panels show the corresponding noisy evolutions. \textbf{(b)} Schematic error budget versus circuit depth. The solid line denotes the total error. The red and cyan dashed lines denote the Trotter error before and after Trotter error mitigation (TEM), respectively. The dark and light gray dashed lines denote the physical noise before and after quantum error mitigation (QEM), respectively. Markers indicate the optimal circuit depth for each case.}
\label{fig:trotternoise_tradeoff}
\end{figure}

\section{Trade-off between Trotter error and physical noise}\label{sec:tradeoff}
In the presence of physical noise, the additional circuit depth required by HOPFs accelerates hardware error accumulation. Fig.~\ref{fig:trotternoise_tradeoff}(a) provides an intuitive illustration of this trade-off relation between Trotter error and circuit depth under a single-qubit Hamiltonian. First, in the noiseless case, the sequential time-evolution trajectory generated by the different sub-Hamiltonians (blue and red lines) deviates from the exact time-evolution trajectory (black dashed curve) generated by the full time-evolution operator $U(T)$. Applying the formula with more steps reduces the deviation between the exact final state $\ket{\psi_f}$ and the Trotterized final state $\ket{\psi_f'}$.

In the upper-left panel of Figure~\ref{fig:trotternoise_tradeoff}, corresponding to the 1-step noiseless (ideal) case, the time step is large and the exact trajectory is approximated rather coarsely. As a result, $\ket{\psi_f'}$ deviates significantly from $\ket{\psi_f}$, indicating a large Trotter error. In contrast, in the upper-right panel, corresponding to the 5-step noiseless case, the time evolution is divided into smaller time steps. The Trotterized trajectory then follows the exact trajectory more accurately, and $\ket{\psi_f'}$ becomes closer to $\ket{\psi_f}$. Thus, in the noiseless case, increasing the circuit depth naturally reduces the Trotter error.

However, when physical noise is included, this behavior is no longer monotonic. In the lower-left panel of Fig.~\ref{fig:trotternoise_tradeoff}(a), corresponding to the 1-step noisy case, the Trotter error remains large, while the accumulated noise is relatively weak due to the lower number of operations. Consequently, the final states obtained from repeated noisy evolutions are distributed within a relatively narrow region. In contrast, in the lower-right panel of Fig.~\ref{fig:trotternoise_tradeoff}(a), despite the reduced Trotter error, the larger number of operations leads to stronger noise accumulation, causing the final states obtained from repeated noisy evolutions to spread over a wider region. 

\begin{table}[t!]
\centering
\begin{tabular}{lcc}
\toprule
Method & $n$ & $q$ \\
\midrule
Leapfrog \cite{strang1968} & 2 & 1 \\
Forest-Ruth \cite{forestruth1990} & 4 & 3 \\
Omelyan \cite{omelyan2003} & 4 & 4 \\
Suzuki \cite{suzuki1990} & 4 & 5 \\
Ostmeyer \cite{ostmeyer2023} & 4 & 5 \\
MO-4th \cite{malezic2026} & 4 & 6 \\
Yoshida \cite{yoshida1990} & 6 & 7 \\
MO-6th \cite{malezic2026} & 6 & 14 \\
\bottomrule
\end{tabular}
\caption{\textbf{Product-formula decompositions used in the benchmark.} $n$ is the leading order of the observable Trotter error in Eq.~\eqref{eq:obserrorexpansion}, and $q$ is the number of two-qubit gate layers per Trotter step, normalized so that Leapfrog has $q=1$. Thus $q=3$ means three times the per-step two-qubit gate depth of Leapfrog.}
\label{tab:integrators}
\end{table}

This trade-off is summarized schematically in Fig.~\ref{fig:trotternoise_tradeoff}(b). While the Trotter error decreases with increasing circuit depth, physical noise accumulates as more operations are applied. The competition between the two renders the total error non-monotonic in depth with a lower bound on the achievable total error, where the optimal circuit depth is set by the physical noise level.

The total error can be further reduced by introducing Trotter error mitigation, which suppresses the algorithmic error without substantially increasing the circuit depth, or by applying quantum error mitigation, which reduces the impact of physical noise at a fixed depth. However, as illustrated in the small-depth regime of Fig.~\ref{fig:trotternoise_tradeoff}(b), if the circuit depth is too small, the time step $\Delta t$ may not be sufficiently small for the perturbative expansion of the Trotter error to be applied to mitigation. In this coarse-grained regime (where the total time $T$ is such that $(\Delta t)^n \sim 1$), removing only a few low-order error terms may not lead to a significant reduction of the total Trotter error.

\begin{figure*}[t]
\centering
\includegraphics[width=\textwidth]{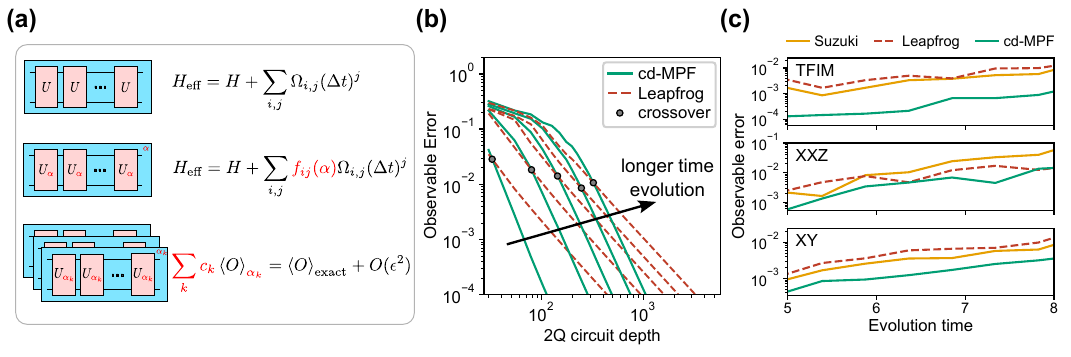}
\caption{\textbf{Constant-depth Multi-Product Formula (cd-MPF).} \textbf{(a)} Schematic of the cd-MPF scheme. cd-MPF augments a Trotter circuit with an auxiliary parameter $\alpha$ and forms a weighted linear combination of the observables measured at several values of $\alpha$, yielding an error-mitigated estimate at constant circuit depth (see method section for details.) \textbf{(b)} Local-observable error (site-averaged $\langle X \rangle$ and $\langle Z \rangle$) as a function of the available circuit depth, comparing cd-MPF (green solid lines) with the second-order Trotter (Leapfrog) formula (red dashed lines). Each line corresponds, from left to right, to evolution times $T = 1, 2, 3, 4$, and $5$. For each $T$, the marker indicates the depth above which the mitigated cd-MPF curve remains below the Leapfrog error. \textbf{(c)} Local-observable error (site-averaged $\langle X \rangle$ and $\langle Z \rangle$) at a fixed circuit depth ($\sim 100$) as a function of evolution time across different spin models, comparing cd-MPF (green solid line), Leapfrog (red dashed line), and the fourth-order Suzuki--Trotter formula (yellow solid line).}
\label{fig:cd-MPF}
\end{figure*}

Throughout this work, we quantify the circuit cost by the two-qubit circuit depth $d$, defined as the number of two-qubit gate layers in the compiled circuit. On current hardware the two-qubit entangling gate is the dominant per-gate source of physical noise, so $d$ serves as a proxy for the physical-noise budget (see
Sec.~S1\,A of the Supplementary Material). For a fixed total evolution time $T$, the number of Trotter steps grows linearly with the depth, $r\propto d$, so the time step $\Delta t=T/r$ scales as $d^{-1}$ and the leading observable error of the Leapfrog formula in Eq.~\eqref{eq:obserrorexpansion} scales as $d^{-2}$.

Under a fixed-depth constraint, a HOPF necessarily operates at a larger effective time step than Leapfrog: a single step of an $n$-th order formula requires $q$ times as many two-qubit gate layers as a single Leapfrog step. The HOPF in this work is benchmarked for $q = 3$--$14$ (see Table~\ref{tab:integrators}). Our depth-matched benchmarks show that this inflation of circuit depth, together with the enlarged step coefficients forced by the Sheng--Suzuki theorem~\cite{Suzuki1991, Wiebe2010}, confines the advantage of the HOPF to absolute Trotter errors below the ${\sim}10^{-2}$ scale, and that even this narrow window effectively vanishes under a depolarizing noise of rate $p = 10^{-5}$, well below current hardware noise levels. (The detailed benchmarks are presented in
Secs.~\ref{sec:comparison} and \ref{sec:noise}, and in
Figs.~S1--S4 of the Supplementary Material.)

\section{Constant-depth multi-product formula}\label{sec:mitigation}

We develop the constant-depth multi-product formula (cd-MPF), which cancels the leading Trotter error at exactly the same two-qubit depth as the underlying Leapfrog circuit. Conventional MPFs~\cite{CarreraVazquez2023, Zhuk2024, Robertson2025} also rely on a classical combination of Trotter circuits, but their combination spans a range of time steps $\Delta t$, so the deepest circuit inevitably grows beyond the single-formula baseline, reintroducing the depth--noise trade-off. 

\begin{figure*}[t]
\centering
\includegraphics[width=1.8\columnwidth]{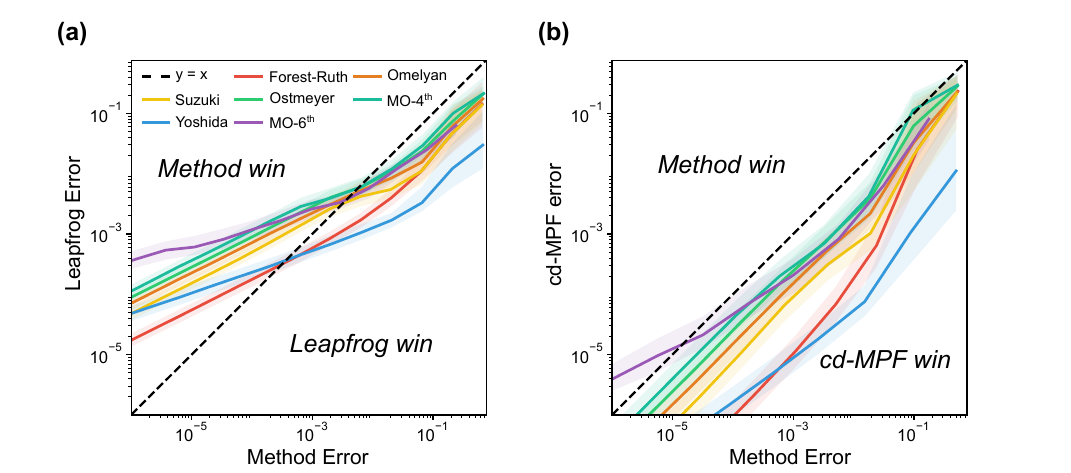}
\caption{\textbf{Higher-order product formulas at matched two-qubit circuit depth.} \textbf{(a)} Local-observable error of each higher-order product formula in Table~\ref{tab:integrators} versus the error of the depth-matched Leapfrog formula, pooled over six spin models and several initial states. Points above the dashed ($y=x$) line mark the regime in which the higher-order formula is more accurate than Leapfrog at the same two-qubit circuit depth; points below mark the opposite. \textbf{(b)} The same higher-order errors versus the cd-MPF error at matched two-qubit depth (with cd-MPF parameter $\gamma\approx2$, see Methods). Points below $y=x$ mark the regime in which cd-MPF is more accurate than the higher-order formula.}
\label{fig:Errorcomparison_HOPF}
\end{figure*}

Instead of spanning different time steps, the cd-MPF combines circuits that share the same gate structure and differ only in a classical parameter~\cite{Lee2025}, a single $\alpha$ that labels circuits carrying different Trotter errors [Fig.~\ref{fig:cd-MPF}(a)]. The building block of cd-MPF is obtained by unsymmetrizing the Leapfrog step. $\alpha$ splits the symmetric $H_A$ half-rotations into unequal fractions,
\begin{equation}
V_\alpha(\Delta t)=e^{-iH_A\alpha\Delta t}\,e^{-iH_B\Delta t}\,e^{-iH_A(1-\alpha)\Delta t},
\end{equation}
while keeping the total per-step $H_A$ rotation angle fixed. Because the net rotation is independent of $\alpha$, the parameter leaves the ideal target evolution untouched and acts only on the Trotter error; to recover the palindromic structure, we combine two such steps into the effective two-step operator 
\begin{equation}
U_\alpha=V_\alpha(\Delta t)\,V_{1-\alpha}(\Delta t).
\end{equation}
The resulting total two-qubit circuit depth is equivalent to that of Leapfrog while the parameter $\alpha$ tunes only the Trotter error. The leading algorithmic error of $U_\alpha$, at order $(\Delta t)^2$, collapses onto only two independent polynomials in $\alpha$, regardless of the number of non-commuting terms in $H$ (see Methods). Measuring the observable at two values $\{\alpha_1,\alpha_2\}$ and forming the linear combination $\sum_k c_k\langle O\rangle_{\alpha_k}$ subject to
\begin{equation}
\sum_{k=1}^{2}c_k=1,\qquad \sum_{k=1}^{2}c_k\,f_i(\alpha_k)=0
\label{eq:cancel}
\end{equation}
removes the $(\Delta t)^2$ term entirely, while every circuit in the combination runs at the same $\Delta t$ and hence the same depth. cd-MPF thus steepens the observable Trotter scaling from $d^{-2}$ to $d^{-4}$ at constant two-qubit circuit depth; the full construction is given in Methods. Solving Eq.~\eqref{eq:cancel} leaves one residual degree of freedom, which we parametrize by the noise-amplification factor
\begin{equation}
\gamma\equiv\sum_k|c_k|;
\end{equation}
$\gamma$ controls the residual $(\Delta t)^4$ error and linearly amplifies uncorrelated noise in the combined estimate, and we use $\gamma\approx2.0$ throughout. 

Fig.~\ref{fig:cd-MPF}(b) benchmarks the scaling behavior of cd-MPF in comparison to the $d^{-2}$ scaling of the depth-matched Leapfrog formula. We observe that the cd-MPF error (green) drops well below Leapfrog (red dashed) and follows the expected $d^{-4}$ slope, confirming the removal of the leading $(\Delta t)^2$ error. Furthermore, the crossover depth, marked for each curve, shifts to deeper circuits as the evolution time $T$ increases. At a fixed depth, a longer $T$ inflates the time step $\Delta t = T/r$, so a deeper circuit is required for the leading-order error to dominate. Below the crossover, the time step is too large for the $(\Delta t)^2$ term to dominate the error expansion---the perturbative regime in which the mitigation is designed to operate---so removing it barely reduces the total error, while the combination weights ($\gamma \approx 2$) amplify the uncancelled residual; in this regime cd-MPF can be less accurate than Leapfrog itself.

However, in contrast to the HOPFs, cd-MPF does not amplify the higher-order error terms. At matched depth, a higher-order formula operates at the inflated time step $\Delta t' = q\,\Delta t$, which magnifies its residual higher-order errors, whereas every circuit in the cd-MPF combination runs at the original $\Delta t$ and leaves the higher-order Trotter error terms essentially unmodified. Within the perturbative regime, cd-MPF therefore attains a lower error than the higher-order formulas at any fixed two-qubit depth, as we show systematically in Sec.~\ref{sec:comparison}.

\section{Benchmarking Trotter formulas}\label{sec:comparison}

Fig.~\ref{fig:cd-MPF}(c) exemplifies the dynamics of the Leapfrog, fourth-order Suzuki~\cite{suzuki1990}, and cd-MPF approaches as a function of evolution time at a fixed two-qubit depth ($\approx100$), a depth accessible on current near-term quantum many-body simulators, for the TFIM, XXZ, and XY models. The evolution-time window is taken where the leading-order error of Leapfrog is sufficiently dominant; at substantially longer times all three methods converge to nearly random, inaccurate results. The fourth-order Suzuki formula (yellow) reaches an inaccurate regime relatively early, whereas cd-MPF (green) tracks the exact dynamics across the window and stays below Leapfrog (red dashed). At equal circuit depth, constant-depth mitigation of the second-order formula is therefore more robust than simply raising the order of the product formula.

We extend the benchmark to the higher-order formulas listed in Table~\ref{tab:integrators}, each characterized by its leading error order $n$ and depth-matching factor $q$, and compare them against Leapfrog and the cd-MPF at the same two-qubit circuit depth. Since HOPFs at a fixed depth operate at the inflated time step $\Delta t'=q\,\Delta t$ (Sec.~\ref{sec:tradeoff}),
Eq.~\eqref{eq:obserrorexpansion} yields the condition for a HOPF to outperform Leapfrog at matched depth,
\begin{align}
\sum_{j\geq2}\omega^{(2)}_{j}\,(\Delta t)^{j} \;>\;
\sum_{j\geq n}\tilde{\omega}^{(n)}_{j}\,(q\Delta t)^{j},
\label{eq:HOvs_Leap}
\end{align}
where $\omega^{(2)}_{j}$ and $\tilde{\omega}^{(n)}_{j}$ denote the observable error coefficients of Leapfrog and of the $n$-th order HOPF, respectively. This condition defines a crossover depth that grows linearly with the target evolution time $T$.

Fig.~\ref{fig:Errorcomparison_HOPF}(a) tests this condition across six spin models at $N=6$ and several initial states (see Methods for the detailed condition). Each higher-order formula beats Leapfrog only once the absolute Trotter error has been driven below $\sim\!10^{-2}$. Above that scale, every higher-order curve falls below the $y=x$ line and becomes \emph{less} accurate than Leapfrog at the same depth, as anticipated by Eq.~\eqref{eq:HOvs_Leap}. The advantages of the HOPFs are therefore confined to a narrow window of small absolute error, reachable only by sufficiently long circuits.

We repeat the same comparison with cd-MPF as the reference [Fig.~\ref{fig:Errorcomparison_HOPF}(b)]. Interestingly, we find a qualitatively different behavior. The advantage window of the cd-MPF spans a much wider error range, with all curves staying below the $y=x$ line from the near-term-relevant scale of $\sim\!10^{-2}$ down well into the perturbative regime below $10^{-5}$. In other words, while a well-tuned higher-order formula does eventually beat Leapfrog once the error is small enough, the cd-MPF beats those same higher-order formulas across both the small-error window in which the HOPFs win against Leapfrog and the larger-error window in which they lose. For any fixed two-qubit depth budget, the cd-MPF therefore makes more efficient use of the circuit than raising the product-formula order does. 

Table~\ref{tab:integrators} reveals the origin of this uniform advantage. The Sheng--Suzuki theorem states that no product formula of order higher than two exists with all-positive coefficients $a_j$ in
Eq.~\eqref{eq:gpf}\cite{Sheng1989, Suzuki1991, GoldmanKaper1996, Blanes2024}.
Cancelling the leading Trotter error requires negative coefficients. For instance, the Forest--Ruth formula ($n=4$,$q=3$) is given as,
\begin{align}
U_{\mathrm{FR}}(\Delta t)=U_{\mathrm{LF}}(b_{1}\Delta t)\,
U_{\mathrm{LF}}(b_{2}\Delta t)\,U_{\mathrm{LF}}(b_{1}\Delta t),
\label{eq:FR}
\end{align}
with $b_{1}\approx1.351$ and
$b_{2}=1-2b_{1}\approx-1.702$. The coefficients satisfy $2b_{1}^{3}+b_{2}^{3}=0$. The backward blocks make each step travel farther than the net time step, enlarging the surviving higher-order error coefficients by orders of magnitude. 

In contrast to the HOPFs, the cancellation of cd-MPF operates on the measured expectation values, and the required negativity is carried by a classical weight [$c_{2}<0$ for $\gamma>1$, Fig.~\ref{fig:cdMPF_condition}(b)], so no block evolves backward and every circuit runs at the original $\Delta t$. The price is a sampling overhead. Estimating the combined observable at a fixed statistical precision requires two circuits and at most a $\gamma^{2}\approx4$-fold increase in the total shot count, independent of the circuit depth. The cd-MPF thus occupies the entry $(n=4,q=1)$ that no unitary product formula can reach.

\begin{figure}[t]
\centering
\includegraphics[width=\columnwidth]{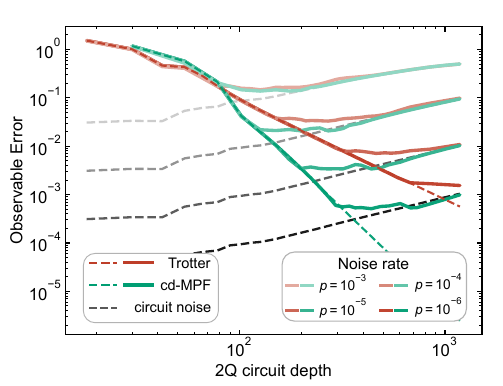}
\caption{\textbf{Crossover between Trotter error and circuit noise.} Total observable error as a function of two-qubit circuit depth at four two-qubit depolarizing rates ($p=10^{-6},10^{-5},10^{-4},10^{-3}$). The dashed red and dashed green curves are, respectively, the noiseless Trotter contribution of Leapfrog and the Trotter contribution of the cd-MPF scheme at $\gamma=2.0$, and the dashed gray curves are the circuit-noise-only contributions at each $p$. The solid red and solid green curves are the total errors without and with mitigation, respectively.}
\label{fig:crossover}
\end{figure}

\section{Optimal circuit depth under explicit noise}\label{sec:noise}

Including an explicit two-qubit depolarizing noise in the depth-matched comparison compresses the already narrow HOPF advantage over Leapfrog further, until it effectively vanishes at any realistic gate-error rate, because the higher-order advantage lives in the large-depth regime where circuit noise is most pronounced (see Fig.~S4 of the Supplementary Material). Only circuits far deeper than those feasible on near-term hardware would render a well-optimized HOPF the more efficient choice.

Figure~\ref{fig:crossover} shows the total observable error versus two-qubit circuit depth, computed by density-matrix simulation at various depolarizing rates, realizing the schematic trade-off of Fig.~\ref{fig:trotternoise_tradeoff}(b). At each $p$ the total error is the sum of a Trotter contribution that decreases with depth ($d^{-2}$ for Leapfrog, $d^{-4}$ for cd-MPF; dashed red and green) and a physical-noise contribution that grows with depth (dashed gray); it is therefore non-monotonic and minimized at an intermediate sweet-spot depth. The steeper $d^{-4}$ suppression of cd-MPF pushes the sweet spot to shallower circuits and lowers the minimum achievable total error across the full range of $p$ examined.

As the figure shows, at $p=10^{-3}$, the noise level typical of present-day commercial quantum computers, cd-MPF and Trotter show little difference, since circuit noise dominates the total error. However, even at $p=10^{-4}$, a level readily attainable in the near term, cd-MPF can already reduce the observable error significantly, and this advantage grows further as the noise rate decreases. Moreover, because cd-MPF mitigates Trotter error at the observable level, its benefit is not tied to a fixed physical noise rate: if the effective noise rate is made sufficiently small---whether by improved hardware or by suppressing circuit noise through quantum error mitigation techniques such as ZNE~\cite{Temme2017, LiBenjamin2017} or PEC~\cite{Temme2017, vandenBerg2023}---the advantage of reducing algorithmic error becomes correspondingly larger.

In conclusion, reducing $p$ alone merely slides the sweet spot along an unchanged Trotter curve, whereas constant-depth Trotter mitigation steepens that curve itself, making it advantageous even for near-term quantum simulation. The two interventions are thus complementary, and the lowest total error is reached only when physical-noise mitigation and constant-depth Trotter mitigation are applied together.

\section{Discussion}

We have analyzed the interplay between algorithmic Trotter error and physical noise in digital quantum simulation under near-term depth constraints, and used it to motivate and characterize a constant-depth Trotter-error mitigation scheme. The depth-matched benchmark of quantum simulation shows that HOPFs do not offer a uniform advantage over the second-order Leapfrog formula: at fixed two-qubit depth their inflated time step and enlarged higher-order error coefficients confine any advantage to absolute errors far below the $10^{-2}$ scale relevant to near-term simulation, and an explicit depolarizing-noise model compresses this narrow window further until it effectively vanishes.

Against this backdrop, cd-MPF provides a practical alternative. A classical linear combination through the tunable splitting parameter $\alpha$ cancels the leading $(\Delta t)^2$ observable error while leaving the two-qubit depth unchanged, steepening the Trotter scaling from $d^{-2}$ to $d^{-4}$. Because the scheme inherits the perturbative boundary of the underlying Leapfrog construction, its breakdown at large $\Delta t$ is gradual rather than catastrophic, and it demonstrates the effectiveness of Trotter-error mitigation for spin-model simulations at a scale accessible on near-term quantum simulators. The steeper suppression provided by cd-MPF, however, materializes only at the deep-circuit end of the trade-off, precisely where physical noise dominates on near-term hardware. As improvements in hardware fidelity, quantum error mitigation, and early fault-tolerant techniques reduce the physical-noise contribution---without eliminating the real cost of circuit depth---the algorithmic side becomes the dominant bottleneck, and constant-depth mitigation of the kind developed here should play an increasingly important role. Constant-depth Trotter mitigation and physical-noise mitigation therefore address opposite sides of the depth--error trade-off, and the lowest total error is reached only when both are deployed together.

A key limitation of expansion-based mitigation, including cd-MPF and MPFs, is that it derives its validity from the Baker--Campbell--Hausdorff (BCH) expansion and becomes unreliable in the large-$\Delta t$ (short-circuit) regime that noisy hardware is often forced to occupy; with only two parameters the present scheme cannot secure an advantage there. Extending constant-depth mitigation to remain effective at large $\Delta t$---for instance by combining additional tunable parameters, problem-specific layer optimization~\cite{tepaske2023optimal, le2025riemannian, mckeever2023classically, mansuroglu2023problemspecific}, or randomized compilation~\cite{childs2019faster, campbell2019random} with classical post-processing---is a promising route to broadening the operating range of useful digital quantum simulation on noisy hardware.

\section*{Acknowledgement}
YK thanks Seung-Woo Lee for early discussions that helped shape the initial direction of this work. SL is supported by a KIAS
Individual Grant via the Quantum Universe Center
(QP104301-6P104301) at Korea Institute for Advanced
Study. This work was supported by the National Research Foundation of Korea (NRF) grant funded by the Korea government (MSIT) (Grants No. RS-2025-25464760, RS-2026-25519864, RS-2025-25446099, RS-2023-NR119928, RS-2025-03392969). This work was also supported by BK21 FOUR (Fostering Outstanding Universities for Research) program through the National Research Foundation (NRF) funded by the Ministry of Education of Korea.

\section{Methods}

\textit{Models and simulation conditions.---}%
This section specifies the spin Hamiltonians and initial states used in the benchmarks of the HOPF and Trotter-error-mitigation performance. $X_{i},Y_{i},Z_{i}$ denote the Pauli operators on site $i$. The spin Hamiltonians used in the benchmarks are three well-known one-dimensional spin models and three random models. The TFIM Hamiltonian is
\begin{align}
H_{\mathrm{TFIM}} = -\sum_{i=1}^{N-1} Z_{i}Z_{i+1} -h\sum_{i=1}^{N} X_{i},
\end{align}
where the Trotter decomposition separates the coupling term ($H_{A}$) from the onsite field term ($H_{B}$). The XXZ Hamiltonian is
\begin{align}
H_{\mathrm{XXZ}} = \sum_{i=1}^{N-1} \bigl( X_{i}X_{i+1}+Y_{i}Y_{i+1}+\Delta\,Z_{i}Z_{i+1} \bigr),
\end{align}
for which the exact coupling on the even and odd bonds is implemented through a KAK decomposition~\cite{vatan2004optimal}, and the Hamiltonian of each bond layer is taken as $H_{A}$ and $H_{B}$, respectively. The XY Hamiltonian is
\begin{align}
H_{\mathrm{XY}} = J\sum_{i=1}^{N-1} \bigl( X_{i}X_{i+1}+Y_{i}Y_{i+1} \bigr),
\end{align}
and, as for the XXZ model, the even-bond and odd-bond couplings are taken as $H_{A}$ and $H_{B}$. These three spin models are all integrable and possess symmetry structures, including translation symmetry, in which case some higher-order terms of the Trotter decomposition can cancel. To exclude such symmetry-induced cancellation and examine the Trotter error for more general Hamiltonians, we also consider a random local spin Hamiltonian, Trotter-decomposed by splitting the even and odd bonds into separate layers; in each layer a distinct random Hermitian, traceless two-qubit generator is assigned to every nearest-neighbor bond, and the corresponding $\mathrm{SU}(4)$ operator is KAK-decomposed for the time evolution. In the main text we use the results of three random Hamiltonians obtained with three different random seeds.

The Trotter error depends not only on the symmetry structure of the Hamiltonian but also on the energy and symmetry of the initial state. For example, when the initial state has sufficiently low energy the Trotter error is small, whereas starting from a high-energy excited state tends to give a relatively large Trotter error. As for the Hamiltonians, to exclude the influence of an individual state we use time-evolution results for a variety of initial states: the N\'eel state $\ket{\psi_{\mathrm{N\acute eel}}}=\ket{0101\cdots01}$, the domain-wall state $\ket{\psi_{\mathrm{DW}}}=\ket{\,\underbrace{0\cdots0}_{N/2}\,\underbrace{1\cdots1}_{N/2}\,}$, the random product state $\ket{\psi_{\mathrm{rand}}}=\bigotimes_{i=1}^{N}\ket{\phi_{i}}$ (each $\ket{\phi_{i}}$ drawn with a fixed random seed), and the ground state of each Hamiltonian. Of these, the noisy density-matrix simulations use only the product states, which are easy to implement as circuits, excluding the ground state. All simulations were performed at $N=6$; given the locality of the Trotter error, however, the results apply to an arbitrary site length, and the site-number dependence of the Trotter error is provided in the Supplementary Material.

\textit{Derivation of cd-MPF.---}%
First, we consider the two-component Hamiltonian $H=H_A+H_B$ with $[H_A,H_B]\neq0$ (we later extend the result to general Hamiltonians which have $K$ non-commuting terms). The $\alpha$-modified step $V_\alpha$ and its palindromic two-step composite are
\begin{align}
V_\alpha(\Delta t)&=e^{-iH_A\alpha\Delta t}\,e^{-iH_B\Delta t}\,e^{-iH_A(1-\alpha)\Delta t},\\
U_\alpha(\Delta t)&=V_\alpha(\Delta t)\,V_{1-\alpha}(\Delta t).
\label{eq:V2alpha}
\end{align}
The full evolution over total time $T$ is implemented as $\bigl(U_\alpha(\Delta t)\bigr)^{r/2}$ with $r=T/\Delta t$. The total $H_A$ rotation per step, $\alpha\Delta t+(1-\alpha)\Delta t=\Delta t$, is independent of $\alpha$, and at $\alpha=1/2$, $U_\alpha$ reduces to two Leapfrog steps, so $U_\alpha$ has the same two-qubit depth as Leapfrog for all $\alpha$; the parameter $\alpha$ is thus a depth-free hyperparameter for the Trotter error. A BCH expansion yields the effective Hamiltonian
\begin{align}
U_\alpha(\Delta t)=&\exp\!\left(-i(2\Delta t)H_{\mathrm{eff}}\right)\nonumber\\
=&\exp\!\Big[-i(2\Delta t)\Big(H+\!\!\sum_{i,\,j\geq2}\! f_{i,j}(\alpha)\,\Omega_{i,j}\,(\Delta t)^{j}\Big)\Big],
\label{eq:Ualpha_BCH}
\end{align}
where $H_{\mathrm{eff}}=H+H_{\mathrm{err}}$; the $\Omega_{i,j}$ are nested commutators of $\{H_\mu\}$ and the $f_{i,j}(\alpha)$ are scalar polynomials. The palindromic structure removes all odd-$j$ terms by symmetry, so the surviving error appears at $(\Delta t)^{2},(\Delta t)^{4},\ldots$, with the leading correction at $j=2$.

At the leading surviving order $j=2$, evaluating the BCH expansion of $U_\alpha$ for $H=H_A+H_B$ yields only two independent polynomial structures in $\alpha$, regardless of the number of non-commuting components of $H$. The reason is that, in $U_\alpha$, all components enter only as the two exponential blocks $H_A$ and $H_B$---each block $e^{-ia_k\Delta t H_A}$, $e^{-ib_k\Delta t H_B}$ is exact on its own, so commutators internal to a block do not contribute to the error---and hence the letters appearing in the BCH expansion are always of just two kinds. The free Lie algebra of length-$3$ nested commutators on two letters has only two basis elements, $[H_A,[H_A,H_B]]$ and $[H_B,[H_B,H_A]]$, whose coefficients are the $\alpha$-polynomials
\begin{align}
f_{1,2}(\alpha) &= \alpha^{2} - 3\alpha + 1 , \\
f_{2,2}(\alpha) &= \alpha - \tfrac{1}{3} .
\end{align}
Since exactly two structures survive at this order, for the remainder of this section we abbreviate $f_i(\alpha)\equiv f_{i,2}(\alpha)$ for $i=1,2$, reserving the double-index notation $f_{i,j}$ for a general order $j$.

The same conclusion holds when the Hamiltonian is split into $K$ separate non-commuting components. For $H=\sum_{\mu=1}^{K}H_\mu$ the palindromic construction extends to
\begin{align}
U_\alpha(t)=\;&\prod_{\mu=1}^{K}e^{-iH_{\mu}(\alpha t)}\prod_{\mu=K}^{1}e^{-iH_{\mu}(\bar\alpha t)}\nonumber\\
&\times\prod_{\mu=1}^{K}e^{-iH_{\mu}(\bar\alpha t)}\prod_{\mu=K}^{1}e^{-iH_{\mu}(\alpha t)},
\end{align}
where $\bar\alpha=1-\alpha$ is the palindromic partner of $\alpha$; time-reversal symmetry again cancels the odd-order terms, leaving the $j=2$ term
\begin{align}
H_{\mathrm{err}}^{(2)}(\alpha)=f_{1}(\alpha)\,C_{1}+f_{2}(\alpha)\,C_{2} \label{eq:Herr2}
\end{align}
as the leading correction, where $C_1$ and $C_2$ denote the operators obtained by summing all length-$3$ nested commutators of $\{H_\mu\}$ that carry the coefficient $f_1(\alpha)$ and $f_2(\alpha)$, respectively---that is, $C_1$ and $C_2$ are themselves sums of nested commutators, generalizing $[H_A,[H_A,H_B]]$ and $[H_B,[H_B,H_A]]$ to $K$ components. One structural fact is central: the $\alpha$-coefficient of a length-$3$ commutator in $H_{\mathrm{err}}^{(2)}$ does not depend on which specific component $H_\mu$ occupies each slot, but only on the repetition pattern among the three slots---that is, whether all three slots hold the same component, exactly two of the three slots coincide, or all three slots are occupied by distinct components---since the palindromic coefficient structure enters identically for every $H_\mu$. On this basis one can show that, for any $K$, $H_{\mathrm{err}}^{(2)}(\alpha)$ retains exactly the same form as Eq.~\eqref{eq:Herr2}, i.e., a linear combination of the same two polynomials $f_1(\alpha)$ and $f_2(\alpha)$. The case $K = 3$ is verified by direct computation: expanding the exact BCH series shows that the coefficients of all length-$3$ commutators built from three components reduce to $f_{1}$, $f_{2}$, or linear combinations thereof---even fully mixed terms such as $[[A,B],C]$ and $[[B,C],A]$ become constant multiples of $f_{1}$ and $f_{2}$.

The general case follows by induction on $K$: since the nested commutator has the fixed length $3$, a newly added component $H_{K+1}$ can occupy $0$, $1$, $2$, or $3$ of the three slots. If it does not appear, the term coincides with a $K$-component term and is a linear combination of $f_{1}$ and $f_{2}$ by hypothesis; if it occupies exactly one slot, only three distinct components are involved, so the term has the same coefficient structure as the $K = 3$ case; if it occupies two slots, only two distinct components appear, again giving a linear combination of $f_{1}$ and $f_{2}$; and if it occupies all three, the commutator $[H_{K+1},[H_{K+1},H_{K+1}]]$ vanishes. Because the length is $3$, these cases are exhaustive, and none produces a polynomial structure outside $f_{1}, f_{2}$. The two leading-order cancellation conditions are therefore the same whether $H$ is split into two blocks or into $K$ components, and the same cd-MPF construction can be applied without modification when $H$ is split into $K$ components.

\begin{figure}[t!]
\centering
\includegraphics[width=\columnwidth]{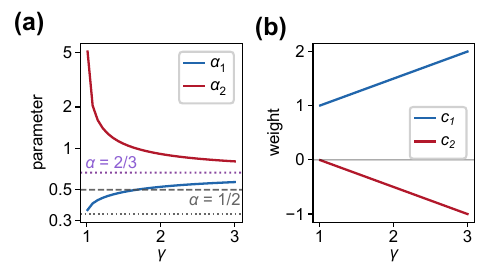}
\caption{\textbf{Parameter and weight for cd-MPF} \textbf{(a)} Values of the two parameters used in cd-MPF as a function of the noise amplification factor $\gamma$. $\alpha=0.5$ coincides with Leapfrog, and the two $\alpha$ values converge to $\alpha=2/3$ as $\gamma\to\infty$. \textbf{(b)} The two weights as a function of $\gamma$.}
\label{fig:cdMPF_condition}
\end{figure}

Because no single $\alpha$ cancels both polynomials, we cancel the leading correction at the level of an \emph{observable} by combining expectation values measured at several values $\{\alpha_k\}$ with weights $\{c_k\}$. Writing $U_\alpha=\exp[-iH(2\Delta t)+\epsilon\sum_i f_i(\alpha)C_i]$ (absorbing $O((\Delta t)^4)$ and higher into $O(\epsilon^2)$) and using the integral identity for the derivative of a matrix exponential,
\begin{align}
e^{A+\epsilon B}=e^{A}+\epsilon\int_{0}^{1}e^{(1-s)A}\,B\,e^{sA}\,ds+O(\epsilon^{2}),
\end{align}
with $A=-iH(2\Delta t)$ and $B=\sum_i f_i(\alpha)C_i$, gives $U_\alpha=U_0+\epsilon\sum_i f_i(\alpha)V_i+O(\epsilon^2)$, where $U_0=e^{-iH(2\Delta t)}$ and $V_i=\int_0^1 e^{-i(1-s)H(2\Delta t)}C_i e^{-isH(2\Delta t)}\,ds$. Inserting this into the expectation value gives
\begin{align}
\langle O\rangle_{\alpha}
=\, \langle O\rangle_{\mathrm{exact}}
+ \epsilon\sum_{i} f_{i}(\alpha)\, \mathrm{Tr}\!\left[O\,\chi_{i}\right]
+ O(\epsilon^{2}) ,
\end{align}
with $\chi_{i}\equiv V_{i}\rho_{0}U_{0}^{\dagger}+U_{0}\rho_{0}V_{i}^{\dagger}$, where the first term coincides with the exact expectation value $\langle O\rangle_{\mathrm{exact}}$. For a weighted combination over $\{\alpha_k\}$, the leading-order error is removed when, together with the observable-normalization condition,
\begin{align}
\sum_{k=1}^2 c_{k} = 1 , \qquad \sum_{k=1}^2 c_{k}\,f_{i}(\alpha_{k}) = 0 \quad (i = 1, 2) ,
\label{eq:conditionnormalize}
\end{align}
which requires two distinct $\alpha_k$ values and two weights. The first condition normalizes the combination so that it reproduces the exact target observable to leading order, and the second cancels the two algorithmic-error structures; the construction is, in this sense, a multi-product formula at the level of expectation values, distinguished from conventional MPFs in that every circuit runs at the same $\Delta t$ and hence the same two-qubit circuit depth.

Solving Eq.~\eqref{eq:conditionnormalize} leaves one residual degree of freedom in $(\alpha_1,\alpha_2,c_1,c_2)$, which we parametrize by
\begin{align}
\gamma \equiv \sum_{k} |c_{k}| .
\label{eq:gammadef}
\end{align}
Equation~\eqref{eq:conditionnormalize} already removes the leading Trotter error for any solution, so the leading-order improvement over Leapfrog is guaranteed by construction. What $\gamma$ actually controls is the residual next-to-leading ($j=4$) contribution, and, in addition, it is tied to the amplification of other noise arising from the linear combination. Larger $\gamma$ selects solutions whose next-to-leading BCH residual is smaller and therefore gives cleaner cancellation in the perturbative regime, but the same $\gamma$ linearly amplifies any uncorrelated noise (shot noise and physical noise alike) in the combined estimate. The scheme therefore favors moderate-to-large $\gamma$ when the underlying Trotter error is small, and degrades at large $\gamma$ when the underlying error is itself large. The values of the parameters and weights as a function of $\gamma$ are illustrated schematically in Figure~\ref{fig:cdMPF_condition}.

More generally, each $(\Delta t)^{j}$ error term in Eq.~\eqref{eq:Ualpha_BCH} carries an $\alpha$-polynomial of degree at most $j$; a single-parameter ansatz can cancel a number of independent polynomial structures bounded by the leading polynomial degree, so the $j=2$ error (exactly two structures) is fully cancellable, but from $j=4$ onward the number of independent structures generically exceeds this bound. Systematic mitigation beyond $j=2$ therefore requires a more general construction with additional tunable parameters, which we leave to future work.

\putbib[references]
\end{bibunit}

\clearpage
\onecolumngrid
\begin{center}
\textbf{\large Supplementary Material}
\end{center}

\begin{bibunit}[apsrev4-2]

\makeatletter
\renewcommand{\@biblabel}[1]{[S#1]}
\makeatother
\renewcommand{\citenumfont}[1]{S#1}
\renewcommand{\bibnumfmt}[1]{[S#1]}

\setcounter{section}{0}
\renewcommand{\thesection}{S\arabic{section}}
\setcounter{figure}{0}
\renewcommand{\thefigure}{S\arabic{figure}}
\setcounter{table}{0}
\renewcommand{\thetable}{S\arabic{table}}
\setcounter{equation}{0}
\renewcommand{\theequation}{S\arabic{equation}}

\setcounter{figure}{0} 
\section{Benchmarking Higher-order product formulas}\label{sec:benchmark}

The main text argues that, under a fixed two-qubit circuit depth constraint, a higher-order product formula (PF) does not uniformly outperform the second-order Leapfrog formula, and that this narrow window of advantage is further compressed by physical noise. This section provides the detailed depth-matched and noisy benchmarks that support that claim, together with a microscopic explanation of the underlying mechanism in terms of Baker--Campbell--Hausdorff (BCH) coefficients.

\subsection{Depth-matched comparison as a proxy for the physical-noise budget}

We compare several higher-order decompositions against the widely used second-order Trotter (Leapfrog) formula, with circuit implementations matched at the same number of parallel two-qubit gate layers. This depth-matched comparison rests on a deliberate simplification: on current superconducting hardware the two-qubit entangling gate is by far the dominant per-gate source of physical noise, so the total physical error of a circuit is set, to leading order, by its two-qubit gate count or equivalently by its two-qubit circuit depth on a fixed connectivity~\cite{Kim2023}. We therefore use two-qubit gate depth as a proxy for the physical-noise budget throughout this section. Later in this section we relax this simplification by re-running the comparison under an explicit two-qubit depolarizing-noise model (Fig.~\ref{fig:noisy_higherorder}) and find that the qualitative picture is preserved.

\subsection{Crossover depth from depth matching}

Under the fixed-depth constraint, higher-order PFs necessarily operate at a larger effective time step than Leapfrog. To see why, consider $H = H_A + H_B$. The elementary repeating block of the Leapfrog scheme is
\begin{align}
U_{\text{LF}}(\Delta t) = e^{-iH_{A}\Delta t/2}\, e^{-iH_{B}\Delta t}\, e^{-iH_{A}\Delta t/2},
\end{align}
which contains three sub-Hamiltonian exponentials. Higher-order PFs are constructed by composing several Leapfrog-like sub-blocks with carefully chosen coefficients, so that one step of an $n$-th order PF contains $q$ such sub-blocks: the standard fourth-order Suzuki construction has $q=5$, and the methods we benchmark span $q$ from $3$ to $14$ (Table I in main text). When such a Trotter step is compiled to hardware, each sub-Hamiltonian exponential costs a fixed number of two-qubit gate layers, so the two-qubit depth contributed by one Trotter step of the higher-order PF is enlarged by the factor $q$ relative to Leapfrog. Consequently, at a fixed total two-qubit depth $d$, the higher-order PF accommodates only $r/q$ Trotter steps where Leapfrog accommodates $r$, and its effective time step is correspondingly inflated from $\Delta t = T/r$ to $\Delta t' = q\,\Delta t = qT/r$.

The condition under which a higher-order PF yields a smaller Trotter error than Leapfrog at matched depth can then be written as
\begin{align}
\sum_{j=2}^{\infty} \omega^{(2)}_{j} \left(\Delta t\right)^{j}
>
\sum_{j=n}^{\infty} \tilde{\omega}^{(n)}_{j} (q\Delta t)^{j} ,
\label{eq:HO_vs_Leap_condition}
\end{align}
where $\omega^{(2)}_{j}$ and $\tilde{\omega}^{(n)}_{j}$ denote the $j$-th order observable error coefficients of the Leapfrog formula and of the $n$-th order higher-order PF, respectively. For example, comparing Leapfrog with the fourth-order Suzuki formula ($q=5$) and retaining the two leading orders of each expansion yields the crossover depth
\begin{align}
d
>
T\sqrt{
5^{4}\,\frac{\tilde{\omega}_{4}^{(4)}}{\omega_{2}^{(2)}}
-
\frac{\omega_{4}^{(2)}}{\omega_{2}^{(2)}}
} ,
\label{eq:crossover_depth}
\end{align}
above which Suzuki outperforms Leapfrog. Since the right-hand side is linear in $T$, the crossover depth grows linearly with the target evolution time. 

\begin{figure}[t]
    \centering
    \includegraphics[width=\textwidth]{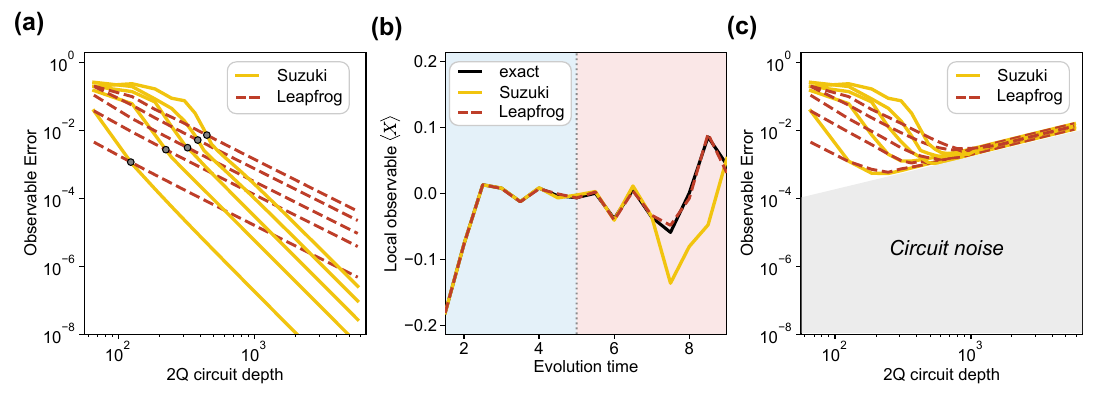}
    \caption{%
    Comparison of the Leapfrog and fourth-order Suzuki product formulas. (a) Site-averaged local-observable error (per-site $\langle Z\rangle$ L1 error, $\mathrm{mean}_i|\langle Z_i\rangle-\langle Z_i\rangle_{\mathrm{exact}}|$, median over four spin models and two initial states; $N=6$) versus two-qubit circuit depth in the noiseless limit, comparing the fourth-order Suzuki product formula (solid yellow) with the second-order Trotter (Leapfrog) formula (dashed red). The five curve pairs correspond to evolution times $T = 1,2,3,4,5$ (larger $T$ gives larger error). For each $T$, the gray circle marks the depth above which the Suzuki curve stays below the Leapfrog error. (b) Site-averaged $\langle X\rangle$ versus evolution time at a fixed two-qubit depth ($\approx 100$; Suzuki $r=10$) for a representative random model, comparing the exact value (solid black) with the Suzuki (solid yellow) and Leapfrog (dashed red) over $T\in[1.5, 9]$. Shading indicates the more accurate formula in each regime (blue: Suzuki; pink: Leapfrog). (c) The same Suzuki--Leapfrog comparison as in (a), with a two-qubit depolarizing error rate $p=10^{-5}$ applied to every two-qubit gate. The gray region marks the error range that circuit noise prevents the curves from reaching at that depth.}
    \label{figS:bench_suzuki}
\end{figure}

This trend is visible in Fig.~\ref{figS:bench_suzuki}(a), which compares the local-observable errors of the fourth-order Suzuki formula ($q=5$) and the depth-matched Leapfrog formula as a function of the two-qubit circuit depth $d$, pooled over spin models and initial states (see the Methods section in the main text for explicit Hamiltonians and initial states used). Yellow solid curves are the Suzuki error and red dashed curves are the Leapfrog error, drawn for evolution times $T= 1,2,3,4,5$ distinguished by curve opacity. For each $T$, the gray circle marks the crossover depth above which Suzuki remains below Leapfrog, and, consistent with Eq.~\eqref{eq:crossover_depth}, these crossover depths shift linearly toward larger values as $T$ increases. Below the crossover, the Suzuki error grows more rapidly than the Leapfrog error, so the higher-order formula is in fact \emph{less} accurate than Leapfrog at the same depth.

Figure~\ref{figS:bench_suzuki}(b) recasts the same information in the time domain, plotting the site-averaged observable $\langle X\rangle$ as a function of $T$ at a fixed two-qubit depth of approximately $100$. The blue and pink shaded regions indicate, respectively, the time windows in which Suzuki and Leapfrog are more accurate, separated by the crossover time inherited from Fig.~\ref{figS:bench_suzuki}(a). In most of the window where Suzuki is superior, both Suzuki and Leapfrog already reproduce the exact dynamics with negligible error. In contrast, past the crossover, Suzuki reaches an inaccurate regime substantially earlier than Leapfrog. Even in the regime where a higher-order PF is advantageous, therefore, the corresponding errors of both formulas are already sufficiently small, and the practical benefit of going to a higher order can be limited.

Figure~\ref{figS:bench_suzuki}(c) repeats the Suzuki--Leapfrog comparison with an explicit two-qubit depolarizing error of rate $p=10^{-5}$ applied to every two-qubit gate. The gray shading marks the error floor set by circuit noise: below this floor the noiseless curves of panel (a) are inaccessible because the physical error dominates the total error. This floor sits precisely inside the depth regime in which the noiseless Suzuki advantage would appear, so under noise the region of genuine Suzuki advantage shrinks substantially even at this modest $p$.

\begin{figure}[t]
    \centering
    \includegraphics[width=\textwidth]{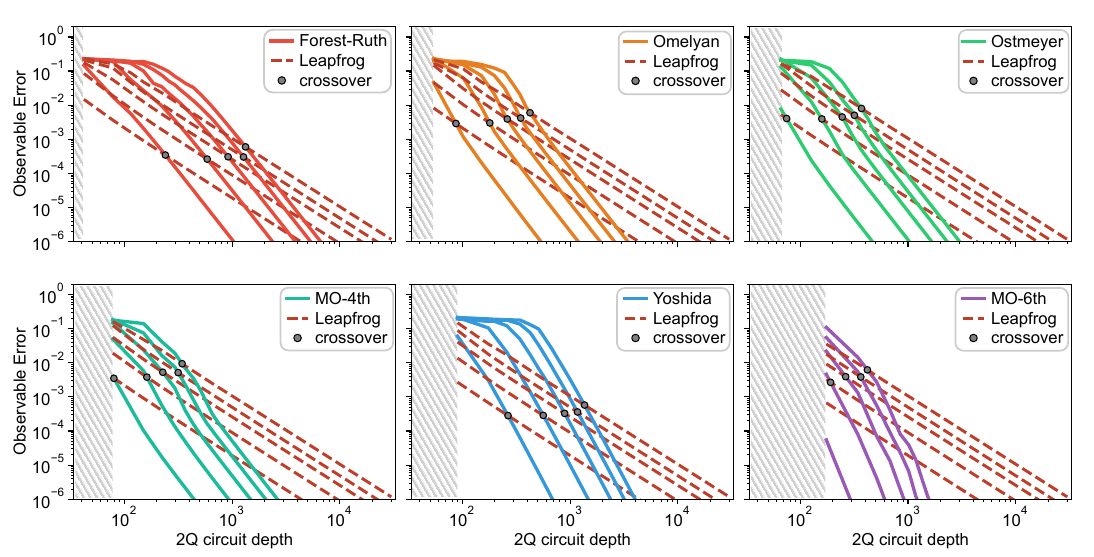}
    \caption{\textbf{Comparison of higher-order product formulas with depth-matched Leapfrog formulas.} As in Fig.~\ref{figS:bench_suzuki}(a), each panel compares one higher-order product formula (solid) with the second-order Trotter (Leapfrog) formula (red dashed) evaluated at the same two-qubit circuit depth. The vertical axis is the site-averaged local-observable error (per-site $\langle X_i\rangle$ and $\langle Z_i\rangle$ error, median over six spin models and five initial states; $N=6$), and the horizontal axis is the two-qubit circuit depth. Within each panel, the five curve pairs correspond to evolution times $T=1,2,3,4,5$ (larger $T$ gives larger error). For each $T$, the gray circle marks the crossover depth at which the method curve drops below the Leapfrog error. The gray dashed region indicates the range that is unreachable due to each method's minimum layer count.}
\label{figS:bench_higher_order}
\end{figure}

To check whether the trend observed for Suzuki persists across a broader class of higher-order formulas, Figure~\ref{figS:bench_higher_order} extends the same depth-matched comparison to all seven higher-order PFs of Table I in the main text. Each panel plots the site-averaged local-observable error of one higher-order method (solid) against the depth-matched Leapfrog error (red dashed), pooled over six spin models (TFIM, XXZ, XY, and three random local Hamiltonians) and five initial states.

Across all seven methods the picture is the same as for Suzuki: at large depth every higher-order PF eventually beats Leapfrog, but this advantage is confined to a window of very small errors. Once the two-qubit depth is reduced below the crossover (gray circles), the higher-order curve rises above the Leapfrog curve, and the same behavior appears for every method and every $T$. Increasing $T$ shifts the crossover to deeper circuits, as anticipated by Eq.~\eqref{eq:crossover_depth}.

Restating this in absolute terms: once the achievable Trotter error exceeds $\sim 10^{-2}$, every higher-order PF is dominated by the depth-matched Leapfrog. Reaching errors well below $10^{-2}$ generally requires deep circuits, in which physical noise has ample interval to accumulate. The very regime in which the higher-order advantage materializes is therefore the regime in which physical noise is most likely to wash it out.

The rapid error growth of higher-order PFs at small $\Delta t'$ observed in the previous subsections can be made concrete by writing out the BCH effective Hamiltonian generated by a single Trotter step. For a general Hamiltonian $H = \sum_\mu H_\mu$, one step of an $n$-th order PF at the depth-matched comparison is generated by
\begin{align}
H_{\mathrm{eff}}^{(n)}
=
H
+
\sum_{i}\sum_{j\geq n}
\beta^{(n)}_{i,j}\,
\Omega_{i,j}\,
(q\,\Delta t)^{j} ,
\label{eq:Heff_BCH}
\end{align}
where $\Delta t$ is the time step at which Leapfrog operates at the same two-qubit gate depth, $q$ is the depth-matching factor of the $n$-th order PF, and $\Omega_{i,j}$ is a nested commutator of $\{H_\mu\}$ of length $j+1$ that contributes at order $(\Delta t)^{j}$. The index $i$ labels the inequivalent nested-commutator structures at that order, and the scalar coefficients $\beta^{(n)}_{i,j}$ are determined by the step coefficients of the formula. By construction, an $n$-th order PF cancels all BCH contributions at orders below $j = n$, so the sum starts at $j = n$.

\begin{figure}[t]
    \centering
    \includegraphics[width=0.5\textwidth]{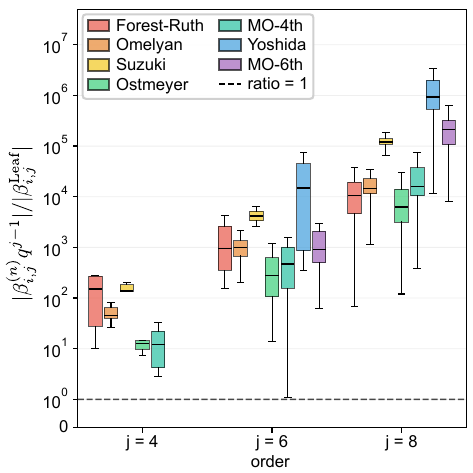}
    \caption{%
    Box plot of the Hall-basis BCH coefficient magnitude ratio between one step of each higher-order method and $q$ consecutive Leapfrog steps at the same two-qubit depth. Each box shows the minimum, maximum, median, and interquartile range over the Hall elements at that order.}
    \label{fig:Hall_coeff_higher_order}
\end{figure}

Equation~\eqref{eq:Heff_BCH} exposes two factors that together govern the higher-order error. The first sits inside the scalar coefficient $\beta^{(n)}_{i,j}$. Cancelling the lower-order Trotter terms that define an $n$-th order PF is achieved by tuning the step coefficients $\{a_{i}\}$ of the formula: the available degrees of freedom suffice to cancel the targeted lower orders but not to suppress the surviving higher-order ones, and for $n\geq 3$ the Sheng--Suzuki theorem forces at least one $a_{i}$ to be negative, which enlarges the magnitudes of the surviving $\beta^{(n)}_{i,j}$~\cite{Suzuki1991, Wiebe2010, Childs2021}. The second factor is the explicit $q^{j}$ accompanying $(\Delta t)^{j}$: at the same two-qubit gate depth, the higher-order PF takes a Trotter step larger than the Leapfrog reference step $\Delta t$ by the depth-matching factor $q$, and this factor enters the BCH expansion raised to the order of each term. The two factors compound multiplicatively, so the effective prefactor controlling the BCH contribution at order $j$ in the depth-matched comparison is $\beta^{(n)}_{i,j}\,q^{j-1}$ rather than $\beta^{(n)}_{i,j}$ alone.

Because $\beta^{(n)}_{i,j}$ follows from the step coefficients of the formula and the commutator algebra of $\{H_\mu\}$ alone, it can be computed analytically once the decomposition $H = \sum_\mu H_\mu$ is fixed, and compared across methods with no reference to the spectrum or initial state of the system. Figure~\ref{fig:Hall_coeff_higher_order} carries out this comparison. For each BCH order $j$, the figure plots the magnitude ratio $|\beta^{(n)}_{i,j}\,q^{j-1}|/|\beta^{\mathrm{Leap}}_{i,j}|$, i.e., the per-step BCH contribution of the higher-order PF including the $q^{j-1}$ inflation from its larger time step, divided by the corresponding contribution of the $q$ consecutive Leapfrog steps that occupy the same two-qubit gate depth. A ratio greater than unity at a given order signals that the higher-order method carries a larger effective prefactor at that order than the depth-matched Leapfrog reference.

As shown in Fig.~\ref{fig:Hall_coeff_higher_order}, these ratios grow rapidly with order, confirming that the two factors combine multiplicatively into a substantially larger effective prefactor than either factor alone would imply. This is directly consistent with Fig.~\ref{figS:bench_suzuki}(a): as the two-qubit depth is reduced, the fourth-order Suzuki formula with $q=5$ reaches an inaccurate regime earlier than Leapfrog because the higher-order BCH terms it carries are inflated by $q^{j-1}$. Among the methods compared, MO-4th and MO-6th, whose step coefficients are explicitly optimized to minimize higher-order BCH contributions, exhibit smaller coefficient ratios than other formulas of the same order --- a direct illustration that the first factor, $\beta^{(n)}_{i,j}$, can be partially suppressed by judicious choice of the step coefficients. In contrast, Yoshida and Forest-Ruth exhibit relatively large ratios, which makes them less efficient than nominally comparable methods in the practical depth regimes considered here.

\subsection{Higher-order PFs under explicit physical noise}\label{subsec:noisy_higher_order}

\begin{figure}[t]
    \centering
    \includegraphics[width=\textwidth]{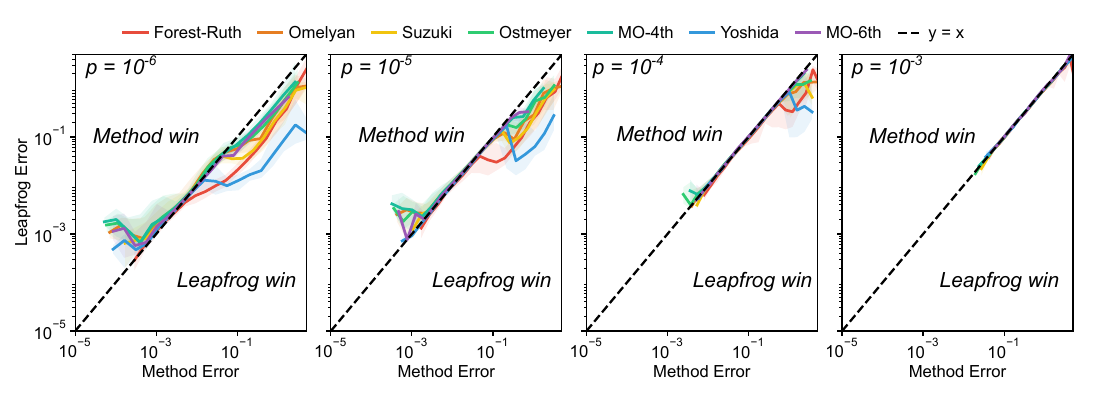}
    \caption{%
    Local-observable error of seven higher-order product formulas versus the depth-matched Leapfrog formula under a two-qubit depolarizing error rate $p$, computed as in Fig.~3(a) of the main text. Each panel fixes one rate $p=10^{-6},10^{-5},10^{-4},10^{-3}$ and shows the total (Trotter + noise) error. Shaded regions indicate the IQR of each plot.}
    \label{fig:noisy_higherorder}
\end{figure}

The final piece of the argument is to check whether the narrow window of higher-order advantage identified in the noiseless benchmarks above survives once physical noise is included. Figure~\ref{fig:noisy_higherorder} repeats the depth-matched comparison of Fig.~\ref{figS:bench_higher_order} via density-matrix simulation at four two-qubit depolarizing rates $p \in \{10^{-6},10^{-5},10^{-4},10^{-3}\}$ applied to every two-qubit gate.

As $p$ increases, the regions in which the higher-order methods previously had an advantage collapse onto the diagonal $y=x$ as they hit the circuit-noise floor. The advantage of the higher-order formulas resides in the small-error regime, which generally requires long circuit depth; circuit noise is pronounced precisely in that long-depth regime and therefore shrinks the region in which higher-order methods retain an advantage. At small depth the Leapfrog advantage survives, but at larger depth the higher-order advantage is washed out. Even at a gate-error rate as small as $p=10^{-5}$ --- already well below what is achievable on near-term hardware --- the higher-order advantage in the small-error regime has effectively disappeared.

Taken together, the range of circuit depths over which higher-order PFs offer a genuine advantage over iterating the second-order Leapfrog is constrained from both sides and is therefore rather limited. At shallow depth, the enlarged higher-order BCH prefactors cause the error to grow more rapidly and Leapfrog prevails; at deep circuits, physical noise dominates and washes out the higher-order advantage. Increasing the target evolution time $T$ shifts the crossover depth to still deeper circuits (Eq.~\eqref{eq:crossover_depth}), further reducing the room to exploit this advantage on near-term hardware. This motivates the question of whether one can suppress the algorithmic error of Leapfrog \emph{without} paying the depth cost of going to a higher order --- the question that the constant-depth multi-product formula (cd-MPF) proposed in the main text is designed to address.

\section{Qubit-number dependence of the Trotter error}\label{app:size_dependence}

The benchmarks in the preceding sections were carried out at $N=6$ sites. What is practically relevant is whether these results extend to larger systems, so in this section we show that the per-site local-observable error is essentially independent of system size $N$ (intensive), and explain the underlying reason following the commutator-scaling framework of Childs et al.~\cite{Childs2021}.

For an $n$-th order product formula $\mathscr{S}_n(\Delta t)$ acting on $H = \sum_{\mu=1}^{K} H_\mu$, Ref.~\cite{Childs2021} established that the per-step error is bounded by
\begin{align}
\left\lVert \mathscr{S}_n(\Delta t) - e^{-iH\Delta t} \right\rVert
\;\lesssim\;
\sum_{\mu_1,\mu_2,\ldots,\mu_{n+1}}
\left\lVert
\bigl[H_{\mu_{n+1}}, \bigl[H_{\mu_n}, \cdots [H_{\mu_2}, H_{\mu_1}] \cdots \bigr]\bigr]
\right\rVert
\,(\Delta t)^{n+1} ,
\label{eq:commutator_scaling}
\end{align}
so the leading per-step error is governed by the norms of length-$(n+1)$ nested commutators of $\{H_\mu\}$. The observable-level error inherits the same structure, and over the total time $T$ the per-step error accumulates over $r = T/\Delta t$ steps to give a total error $\propto T\,(\Delta t)^{n}$.

All spin models we benchmark (TFIM, XXZ, XY, and random local Hamiltonians) contain only nearest-neighbor interactions, so each $H_\mu$ is supported on a single bond layer that acts nontrivially on only two neighboring sites. For any two local terms $H_\mu$ and $H_\nu$, the commutator $[H_\mu, H_\nu]$ is nonvanishing only when their supports overlap and is exactly zero otherwise. Consequently, a length-$(n+1)$ nested commutator
\begin{align}
[H_{\mu_{n+1}}, [H_{\mu_n}, \cdots [H_{\mu_2}, H_{\mu_1}] \cdots]]
\end{align}
is nonzero only when the supports of all constituent terms form a connected chain, and its support is confined to at most $O(n)$ neighboring sites --- a manifestation of the effective light-cone structure imposed by the locality of the interactions~\cite{lieb1972finite}.

The locality of the single-step nested commutators is preserved when many steps are iterated. To leading order, an $r$-fold repetition of the product formula is equivalent to evolving with a single time-independent effective error Hamiltonian $H_{\mathrm{err}}$ for the full duration $T$,
\begin{align}
[\mathscr{S}_n(\Delta t)]^r \;\approx\; \exp\!\left[-i\,T\,(H + H_{\mathrm{err}}(\Delta t))\right] ,
\end{align}

so the operator generating the observable error retains exactly the same local structure as the single-step $H_{\mathrm{err}}$: it is a sum of local terms, each supported on $O(n)$ neighboring sites, and the number of distinct term species is independent of $N$. The error of a local observable $O_i$ is therefore sourced only by those terms in $H_{\mathrm{err}}$ whose supports lie within a finite neighborhood of $O_i$. By the Lieb--Robinson bound, the effective support that can influence $O_i$ over evolution time $T$ grows at most as $O(vT)$, where $v$ is a Lieb--Robinson velocity set by local couplings alone. Crucially, this effective support depends on $T$ but \emph{not} on $N$: once $T$ is fixed, adding sites far outside the light cone of $O_i$ cannot change $\langle O_i\rangle$ or its error. For the fixed evolution times $T\leq 5$ used in our benchmarks, the light cone is bounded by a system-size-independent constant.

Combining the above, the total Trotter-error norm bound scales as
\begin{align}
\left\lVert [\mathscr{S}_n(\Delta t)]^{r} - e^{-iHT} \right\rVert
\;\lesssim\;
N \cdot g_n \cdot (\Delta t)^{n} ,
\label{eq:extensive_bound}
\end{align}
where $g_n$ is a local constant independent of $N$: the total error is extensive in $N$, but the per-site error $N^{-1}\lVert\cdots\rVert$ is $N$-independent at leading order. The same conclusion holds at the observable level: for a local observable $O_i$ (e.g., $\langle X_i\rangle$ or $\langle Z_i\rangle$), the site-averaged error $\mathrm{mean}_i|\langle O_i\rangle - \langle O_i\rangle_{\mathrm{exact}}|$ approaches a finite, $N$-independent value. This locality argument holds cleanly in the perturbative regime, where only the leading few $(\Delta t)^{n}$ terms dominate. Higher-order terms involve longer nested commutators with wider supports, but as long as the leading order dominates, the per-site error is set by essentially local information.

\begin{figure}[t]
    \centering
    \includegraphics[width=\textwidth]{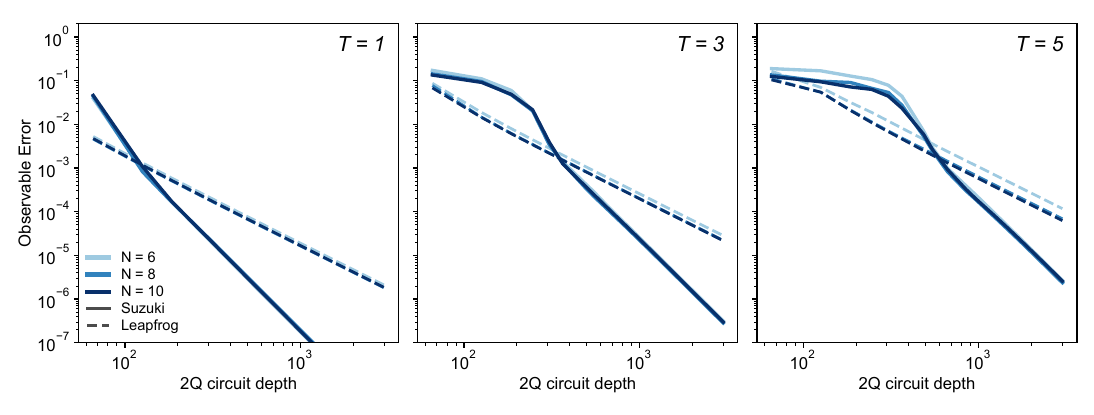}
    \caption{%
    System-size dependence of the Trotter error. Site-averaged local-observable error (median over six spin models and five initial states; noiseless) versus two-qubit circuit depth for the fourth-order Suzuki (solid) and depth-matched Leapfrog (dashed) formulas, at evolution times $T=1,3,5$ (left to right). In each panel the system sizes $N=6,8,10$ are overlaid (color = $N$, light to dark).}
    \label{fig:size_dependence}
\end{figure}

Figure~\ref{fig:size_dependence} directly confirms this prediction. We repeat the fourth-order Suzuki vs.\ depth-matched Leapfrog comparison of Fig.~\ref{figS:bench_suzuki}(a) at three system sizes $N=6,8,10$, at evolution times $T=1,3,5$ (left to right). The curves for different $N$ nearly coincide at every evolution time considered, confirming that the per-site local-observable error is intensive in system size. Consequently, the depth-matched benchmarks reported at $N=6$ throughout this Supplementary Material apply, up to per-site normalization, to larger site counts as well, provided the total evolution time $T$ is kept fixed and the leading-order Trotter contribution dominates.

% \bibliography{references.bib}

\putbib[references]
\end{bibunit}

\end{document}